# Self-induced Faraday instability laser

A. M. Perego[1,*], S. V. Smirnov[2], K. Staliunas[3,4], D. V. Churkin[2], S. Wabnitz[2,5]

[1]Aston Institute of Photonics Technologies, Aston University, Aston Express Way B4 7ET, Biriminghan, United Kingdom
[2]Novosibirsk State University, 1 Pirogova str., Novosibirsk, Russia
[3] Departament de Fisica i Enginyeria Nuclear, Universitat Politècnica de Catalunya, E-08222, Barcelona, Spain
[4]Institució Catalana de Recerca i Estudis Avançats, Passeig Lluis Companys 23, E-08010, Barcelona, Spain
[5]Dipartimento di Ingegneria dell'Informazione, Università di Brescia, and INO-CNR, Via Branze 38, Brescia 25123, Italy

*corresponding author: peregoa@aston.ac.uk

*Abstract*

We predict the onset of self-induced parametric or Faraday instabilities in a laser, spontaneously induced by the presence of pump depletion in the cavity, which leads to a periodic gain landscape for light propagating in the cavity. As a result of the instability, continuous wave oscillation becomes unstable even in the normal dispersion regime of the cavity, and a periodic train of pulses with ultrahigh repetition rate is generated. Application to the case of Raman fiber lasers is described, in good quantitative agreement between our conceptual analysis and numerical modeling.

*Introduction*

Modulation instability (MI) is a universal mechanism that leads to the break-up of continuous waves (CWs) into modulated patterns in weakly dispersive and nonlinear physical systems. MI-activated pattern formation has been observed in a variety of different systems, from hydrodynamics to plasmas and optics [1-3]. Transverse pattern formation in lasers, which has been actively studied since late 80-ies, is also due to MI initiating the instability and light patterns in the transverse direction to optical axis of the laser [4-7]. In the case of single-transverse, multi longitudinal mode lasers, e.g., fiber lasers, MI can occur along the laser cavity, and break the stability of CW oscillation. Hence it may permit the generation of a continuous train of pulses with ultrahigh repetition rates [8], which is of great interest in many applications, from optical metrology to communications. The initial proposal to achieve a MI laser was based on the synchronous and coherent injection of an intense laser beam into a passive optical fiber ring cavity [8-11]. In this way, MI-induced pulse train generation in a nonlinear dispersive cavity results in the formation of a coherent or mode-locked frequency comb. Even in the absence of an external pump laser, for sufficiently high-level pumping of the active medium in the cavity, MI may lead to the spontaneous transition from CW to pulsed operation, leading to the so-called "self-induced MI laser" [12-13]. Stabilization of the laser repetition rate and simultaneous substantial reduction of the laser threshold can be achieved by inserting a frequency-periodic linear filter in the active nonlinear cavity [14-15]. It has been observed that in this case the mechanism leading to pulse train formation can be more conveniently described in terms of a dissipative four-wave mixing process, which leads to self-induced pulse train formation, irrespective of the sign of the cavity group velocity dispersion [16-18].

In this Letter, we propose and theoretically analyze a new mechanism for pulse train and frequency comb generation in lasers, based on parametric (or Faraday) instability. Parametric instabilities (PIs) are a well-known universal mechanism for pattern formation in many different branches of physics [19]. PIs in nonlinear and dispersive wave propagation occur whenever one of the medium parameters is periodically modulated along the longitudinal direction. For externally periodically forced systems, PI is also commonly known as Faraday instability, following the initial Faraday's observation of pattern formation induced by the modulation of the vertical position of an open fluid tank [20]. Besides hydrodynamics, Faraday-like patterns are observed in a wide range of physical settings, from crystallization dynamics, chemical systems, and optics [21-25]. In the context of fiber lasers, it has been proposed that Faraday instability can lead to the generation of pulse trains when the cavity dispersion is periodically modulated [26-30].

In addition to parametrically forced systems, there are many physical systems which feature the presence of collective oscillations, and thus may spontaneously lead to parametric or Faraday instabilities. Consider for example Bose-Einstein condensates, where PI may be introduced by the harmonic modulation of the nonlinear interaction or the profile of the trapping potential [31-33], and nonlinear graded index multimode optical fibers, where a self-induced intensity grating results from beam self-imaging [34]. Here we point out that a laser is a fundamental physical system where a periodic modulation of the gain naturally occurs in the cavity in the presence of pump depletion (See Fig. 1). The resulting Faraday instability is analogous to the PI that has been predicted in periodically amplified fiber optic communication links [35]. However, in a laser the wave dynamics is much more complex, as it requires consideration of the effects of finite gain bandwidth (or temporal diffusion) and nonlinear gain saturation, as it can be described in terms of the Ginzburg-Landau equation [36]. Note that a sideband instability also occurs in soliton lasers, but in that case the generation of sidebands is of a different nature, since it is due to the coupling between the soliton and dispersive waves [37-39].

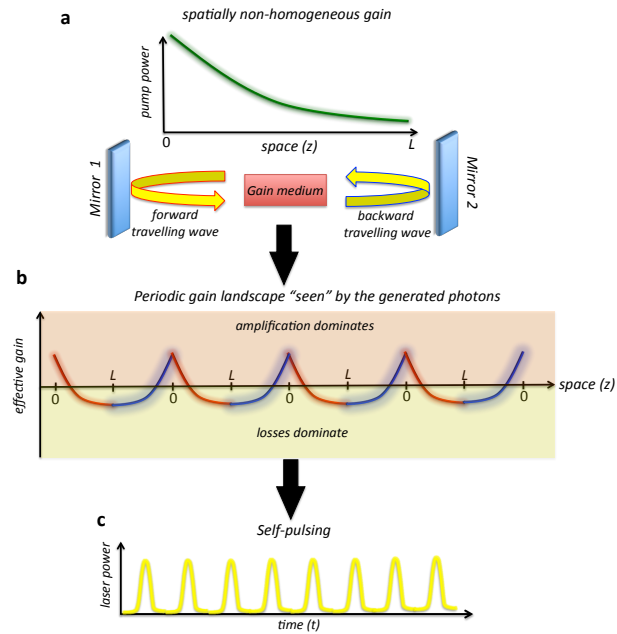

**FIGURE 1** Concept of self-induced Faraday instability laser: The spatially inhomogeneous gain profile naturally arising from the solutions of the nonlinear laser equations **a** gives rise to an effective periodic gain (and consequently nonlinearity) landscape profile **b** seen by the generated photons that travel back and forth in the linear cavity. Such a periodic gain and nonlinearity variations result in a parametric forcing leading to self-pulsing with high repetition rate **c**.

*Self-induced Faraday instability in the CGLE*

As we have already mentioned in the introduction, our goal is to characterize the self-induced Faraday instability where a parametric modulation arises spontaneously from the solutions of a nonlinear system –the laser- and it is not imposed from the external world. However, we first illustrate the generic features of the dynamical instability and of the associated pattern formation in the simplified and idealized case of the complex Ginzburg-Landau equation (CGLE), while in the second part of the article we will provide a specific example based on numerical simulations of a realistic Raman fiber laser.

We have considered the following complex CGLE for the field envelope $A(z,t)$, defined in the local time reference frame $t$ and evolving along the spatial coordinate $z$:

$$\frac{\partial A}{\partial z} = \mu(z)A + (b-id)\frac{\partial^2 A}{\partial t^2} + (ic-s)|A|^2 A. \qquad \text{Eq. 1}$$

The nonlinearity coefficients describe self-phase modulation $c$ and gain saturation $s$, respectively, while $d$ and $b$ describe dispersion and finite gain bandwidth (diffusion). Across our study, we have taken all coefficients with positive values, (Benjamin-Feir stable regime). The spatial dependent gain coefficient mimics the periodic gain profile experienced by the electric field upon propagation. The modulated gain coefficient is $\mu(z)=\mu_{av}+\delta_\mu \cos(k_m z)$ with $\mu_{av}$ equal to the average gain; the modulation occurs with modulation depth $\delta_\mu$ and with spatial period $L_m$ which defines the modulation wave number $k_m=2\pi/L_m$.

Since we are dealing with a periodic system, a Floquet analysis allows us to characterize the instability showing which modes experience amplification due to periodic forcing.

The procedure for the Floquet analysis is as follows: we first calculated numerically the stationary solution of the field spatial distribution by suppressing any temporal modulation. For each mode with frequency $\omega$, we have then computed a 4-by-4 stability matrix whose entries are given as follows: the first and second row entries are the real and imaginary parts of the modes $+\omega$ and $-\omega$ amplitudes after the evolution of respectively real and imaginary perturbations to mode $\omega$. The third and fourth rows of the stability matrix contain the real and imaginary parts of $+\omega$ and $-\omega$ mode amplitudes, respectively, after the evolution of real and imaginary perturbations added to mode $-\omega$. The evolved modes' amplitudes were normalized to the initial perturbation's absolute value. Diagonalizing the stability matrix gives 4 eigenvalues –Floquet multipliers- relative to modes $+\omega$ and $-\omega$.

For each mode, we have defined the instability power gain as $G=2\ln(|\lambda_m|)/L_m$, where $\lambda_m$ is the mode's Floquet multiplier with the largest absolute value. Instability takes place when $G>0$. Since the Floquet spectrum is symmetric with respect to frequency, throughout the paper we have plotted only the positive frequency side.

For large enough $\delta_\mu$, we have observed a destabilization of the CW solution with net growth of modulation modes. A dependence of the instability gain on relevant parameters of the system is summarized in Fig.2 .

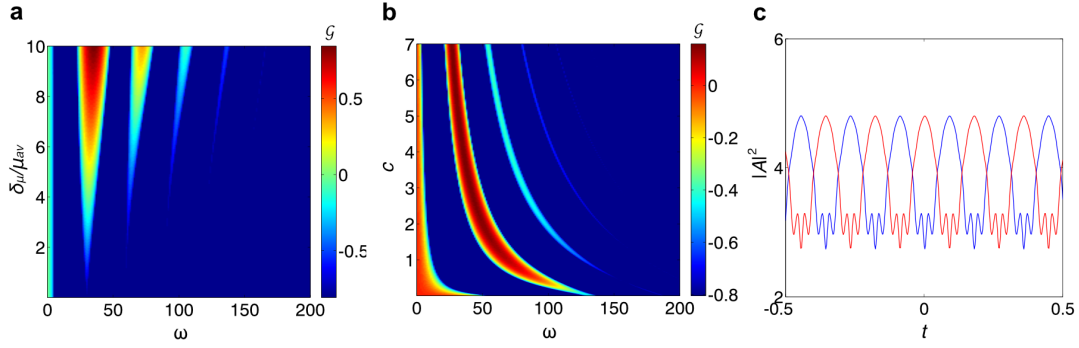

**FIGURE 2** Periodic variation of the gain induces MI when the modulation depth $\delta_\mu$ crosses a certain threshold: MI develops with synchronization area increasing with $\delta_\mu$. In **a,** the instability gain $G$ is plotted in the $\omega$-$\delta_\mu/\mu_{av}$ plane. The instability frequency decreases with nonlinearity $c$, **b**, hence showing that we are in presence of a Faraday instability. Pattern formation: in **c** the stable temporal pattern, i.e., pulses on the finite background, is depicted. Blue and red lines correspond to the field modulus squared plotted at even (blue) an odd (red) modulation periods, respectively. Common parameters to all plots are $c$=5 $s$=0.1 $d$=1.18•10$^{-4}$, $\mu_{av}$=0.4, $\delta_\mu$=5$\mu_{av}$, $L_m$=1.5, $b$=0; except for the case in **a,** where $\delta_\mu$ varies from 0 to 10$\mu_{av}$.

As one can expect, an increase in the forcing strength results in a broader synchronization region, which is typical for parametric resonances. Furthermore, together with the appearance of low frequency instability tongues which dominate the dynamics, we observe the presence of underdeveloped instability tongues ($G$<0), which correspond to higher-order parametric resonances.

A common feature of parametric instabilities in the normal dispersion/diffraction regime is the inverse scaling of the instability frequency versus nonlinearity. This trend is confirmed in our study (see Fig. 2b). The growth of modulation unstable modes eventually results in a stable pattern formation. Here, similarly to other studies on parametric instabilities [29], we observe a period-2 dynamics of the pattern, corresponding to even and odd numbers of the modulation period, respectively (Fig. 2c).

It is indeed not surprising to recover here, at least qualitatively, the typical features of parametric instabilities, which in general are induced by the periodic modulation of dispersion and nonlinearity. In fact, the gain modulation considered automatically affects the nonlinearity, which leads to an effective spatially periodic modulation of the self-phase modulation coefficient $c$.

In the case of parametric instabilities, the first excited temporal mode has angular frequency $\omega_{inst}$ given by:

$$\omega_{inst} \approx \frac{\pi}{L_m \sqrt{2cd\,\mu_{av}/s}}\,. \qquad \text{Eq. 2}$$

The resemblance remains here however on a qualitative level, and Eq. 2 can be just considered as a rule of thumbs, since the more involved nature of the dissipative modulation does not allow to stretch the analogy further.

*Coexistence of spatiotemporal chaos and Faraday patterns*

Achieving pulses which are stable and robust is a highly required feature of mode-locked lasers. Hence, understanding the impact of lasers' parameters on the pulses stability is of paramount importance. In this respect, we have preliminarily investigated the impact of the diffusion coefficient *b* (gain-bandwidth) on the stability of the self-induced Faraday pulses. It is a widespread belief that losses play a stabilizing role in pattern formation processes. Surprisingly enough, as illustrated in Fig.3, we observe that this is not the case for the particular situation that we are considering. Indeed, numerical simulations of Eq.1 in presence of diffusion show that parametric forcing destabilizes the homogeneous solution, leading to the onset of spatiotemporal chaos. First of all, Faraday patterns are excited, however they are dynamically unstable (for the particular choice of the parameters used in Fig. 3a an 3b). As a consequence, these patterns rapidly break-up, leading to a well-known scenario, which is typical for spatiotemporal extended chaotic systems [36,40-42]. Nevertheless, we notice the interesting presence of Faraday pattern "islands" on a "sea" of spatiotemporally chaotic turbulence (See Fig. 3a and 3b). If diffusion is reduced to below $b=10^{-6}$, we surprisingly observe that the laser dynamics is dominated by stable Faraday patterns (Fig. 3c), without any signature of turbulence. These considerations may have a relevant impact towards the practical design of Faraday instability lasers.

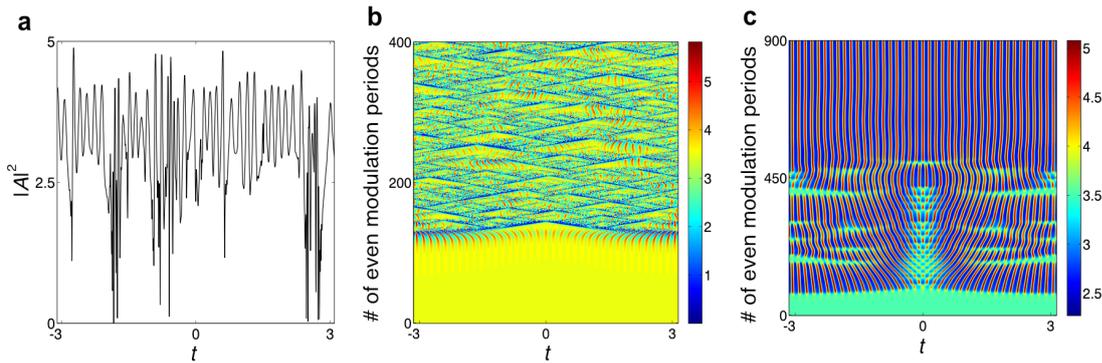

**FIGURE 3** In **a**, we depict a typical example of a turbulent temporal intensity dynamics where periodic oscillations can still be observed. In **b,** the spatiotemporal evolution of $|A|^2$ shows, in the presence of diffusion (*b* different from zero), the coexistence of spatiotemporal chaos and parametric patterns (high intensity red stripes); **a** is a section of **b**. In **c,** by reducing diffusion we obtain a stable pulse train. Parameters used in **a** and **b** are *c*=5 *s*=0.2 *d*=1.18•10$^{-4}$, $\mu_{av}$=0.8, $\delta_\mu$=5$\mu_{av}$, $L_m$=1.5, *b*=1.97•10$^{-5}$. **c** is obtained with the same parameters as **a** but with *b*=0.

*Self-induced Faraday instability in a fiber Raman laser*

As a proof of concept, we show by numerical simulations that self-induced Faraday instability owing to periodic gain variation can indeed be observed in a realistic laser system. We consider here the all-normal dispersion linear cavity Raman fiber laser.
The pump field is injected at one cavity mirror, while the Stokes is generated spontaneously through the Raman scattering process along the fiber. Linear fiber attenuation together with pump depletion lead to a longitudinally dependent pump profile along the fiber (Fig. 4a). The pump "modulation depth" becomes progressively pronounced with increasing values of its power. The pump field is most intense near one cavity mirror, while it is much weaker in the vicinity of the other one. Hence the Stokes field, while travelling back and forth in the cavity, "experiences" both a periodic gain and a periodic nonlinearity profile. When the gain variation is sufficiently large, the threshold for parametric instability is crossed, and the amplification of spectral sidebands takes place. As a consequence, the CW solution of the cavity loses stability. Correspondingly, a pulse train is generated, with a repetition rate dictated by the instability frequency.
Light evolution in the laser is described by a set of four coupled generalized nonlinear Schrödinger equations for the forward and backward propagating fields $A^{\pm}_{p,s}$, respectively, where suffixes $p$ and $s$ refer to pump and Stokes fields

$$\pm \frac{\partial A^{\pm}_p}{\partial z} = -\beta_{1p} \frac{\partial A^{\pm}_p}{\partial t} - i\frac{\beta_{2p}}{2}\frac{\partial^2 A^{\pm}_p}{\partial t^2} - \frac{\alpha_p}{2} A^{\pm}_p + i\gamma_p |A^{\pm}_p|^2 A^{\pm}_p - \frac{g_p}{2}\left(|A^{\pm}_s|^2 + \langle |A^{\mp}_s|^2 \rangle\right) A^{\pm}_p$$

$$\pm \frac{\partial A^{\pm}_s}{\partial z} = -i\frac{\beta_{2s}}{2}\frac{\partial^2 A^{\pm}_s}{\partial t^2} - \frac{\alpha_s}{2} A^{\pm}_s + i\gamma_s |A^{\pm}_s|^2 A^{\pm}_s + \frac{g_s}{2}\left(|A^{\pm}_p|^2 + \langle |A^{\mp}_p|^2 \rangle\right) A^{\pm}_s. \qquad \text{Eqs. 3}$$

Here $\gamma_{p,s}$, $\beta_{2p,s}$, $\alpha_{p,s}$ and $g_{p,s}$ denote the Kerr nonlinearity coefficient, group velocity dispersion, attenuation, and Raman gain for pump and Stokes wavelengths, respectively, while $\beta_{1p}$ describes the group velocity mismatch and the bra-kets temporal average.
The stability properties of the homogeneous solution of the Raman laser can be characterized once again by means of a Floquet linear stability analysis, whose results are depicted in Fig. 4b. In order to perform the Floquet analysis of the Raman laser homogeneous solution, we first computed numerically the spatial profiles of pump and signal fields in the stationary state, by suppressing temporal modulations. After that, we repeated the Floquet sideband analysis as previously described integrating numerically Eqs. 3 over one full cavity round trip.
In Fig. 4b, the Faraday instability spectrum shows the presence of exponentially growing modulation sidebands. Consistently with these theoretical predictions, full numerical simulations of Eqs. 3 reveal that temporal Faraday patterns, consisting in a train of pulses, are indeed generated, owing to the self-induced Faraday instability. For a quantitative comparison with predictions of the linearized stability analysis, the repetition rate of the numerically generated pulse trains measured from the power spectrum peak is indicated by black dots in Fig. 4b. As can be seen, the linear stability analysis provides a good estimation of the pulses repetition rate. The generated pulses are in general regular (see Fig. 4d), but sometimes are subject to transient instabilities and collisions; after those turbulent events the regular pulsation regime regenerates spontaneously (Fig. 4e).

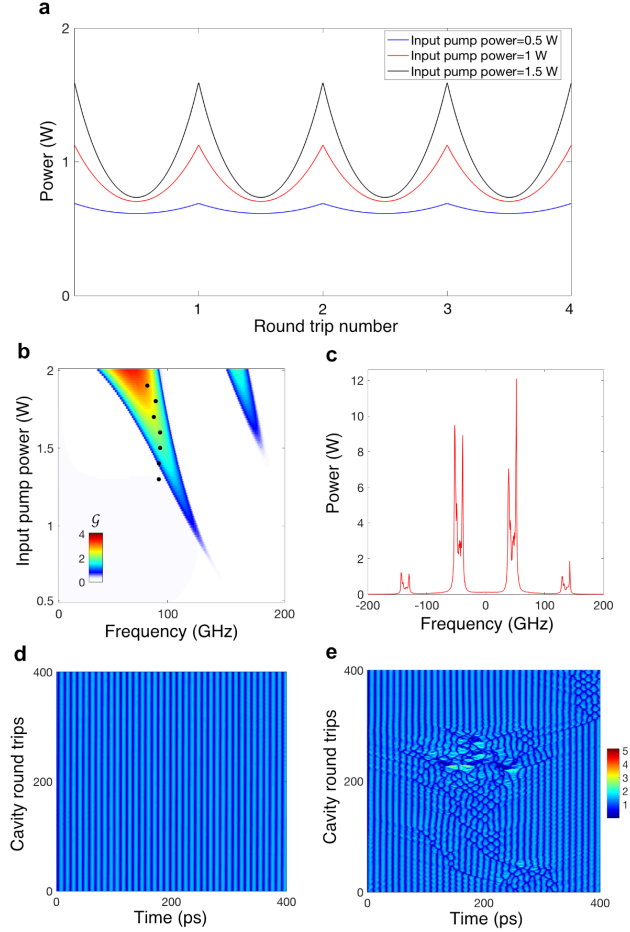

**FIGURE 4** In **a,** the total pump power "experienced" by the Stokes field in the stationary state is plotted versus the longitudinal coordinate, for a length corresponding to four cavity round trips, and for various input pump power values. In **b**, the Faraday instability gain *G* is plotted versus pump power and frequency: colored areas correspond to unstable modes. (black dots denote generated pulses repetition rate). In **c,** an example of optical spectrum is plotted for an input pump power equal to 1.3 W. In **d** and **e,** examples of stable and metastable pulses for input pump power equal to 1.3 W are depicted. Parameters used are $\gamma_p$=3 (W km)$^{-1}$, $\gamma_s$=2.57 (W km)$^{-1}$, $g_p$=1.51 (W km)$^{-1}$, $g_s$=1.3 (W km)$^{-1}$, $\alpha_s$=0.8 km$^{-1}$, $\alpha_p$=0.5 km$^{-1}$, fiber length *L*=0.37 km. At the cavity boundaries, radiation is reflected by super Gaussian (order 3) unchirped fiber Bragg-gratings having 1nm FWHM.

We stress that periodic forcing in this case is not due to any action performed on the system from the external world, but it arises spontaneously, owing to the particular self-organization process of the field solutions, hence the denomination: self-induced Faraday instability. This feature distinguishes our laser system from previous and recent studies of dispersive and dissipative parametric instabilities [21, 23-26, 43], where the forcing is produced by a suitably designed dispersion landscape or by cavity boundary conditions.

*Conclusions*

In conclusion, our study sheds light on an up to now not considered case of self-induced PI, and on the associated pattern formation process, taking place in laser systems described by the universal CGLE, and its more complex versions. We have shown, based on realistic examples, that lasers with a suitable inhomogeneous spatial distribution of gain along the resonator can be the ideal platform for observing the predicted self-induced Faraday instability. The primary interest in the associated temporal pattern formation process based on Raman gain is linked to the possibility of exploiting the self-induced Faraday instability to design pulsed fiber laser sources operating at high repetition rates, widely tunable across the entire transparency window of optical fibers.

# Acknowledgements

This project has received funding from the European Union's Horizon 2020 research and innovation program under the Marie Skłodowska-Curie grant agreement No 691051. Authors acknowledge financial support from Spanish Ministerio de Ciencia e Innovación, and European Union FEDER through project FIS2015-65998-C2-1-P, and partial support of the Turkish Academy of Science. The work of S. W., S. V. S and D. V. C. is supported by Ministry of Education and Science of the Russian Federation (Minobrnauka) (14.Y26.31.0017). A. M. P. acknowledges support from the ICONE Project through Skłodowska-Curie Grant No 608099 .